\documentstyle[Sprocl]{article}
\begin{document}
\def\lambdabar{{\mathchar"26\mkern-9mu\lambda}}
\title{GRAVITOELECTROMAGNETISM}
\author{BAHRAM MASHHOON}
\address{Department of Physics and Astronomy\\
University of Missouri-Columbia\\
Columbia, Missouri 65211, USA\\
E-mail: mashhoonb@missouri.edu}
\maketitle
\abstracts{Gravitoelectromagnetism is briefly reviewed and some recent
developments in this topic are discussed.  The 
stress-energy content of the gravitoelectromagnetic field is described from different standpoints.  In particular, the
gravitational Poynting flux is analyzed and it is shown that there exists a steady flow of gravitational energy circulating
around a rotating mass.}

\section{Introduction}

Gravitoelectromagnetism (GEM) is based upon the close formal analogy between Newton's law of gravitation and Coulomb's law
of electricity.  The Newtonian theory of gravitation may thus be interpreted in terms of a gravitoelectric field.  Any field
theory that would bring Newtonian gravitation and Lorentz invariance together in a consistent framework would necessarily
contain a gravitomagnetic field as well.  In general relativity, the non-Newtonian gravitomagentic field is due to mass
current and has interesting physical properties that are now becoming amenable to experimental observation.

Developments in electrodynamics in the second half of the nineteenth century led Holzm\"{u}ller [1] and Tisserand [2] to postulate a
gravitomagnetic component for the gravitational influence of the Sun on the motion of planets.  The magnitude of this ad hoc component could
be adjusted so as to account for the excess perihelion motion of Mercury.  However, Einstein's general relativity successfully accounted for
the perihelion precession of Mercury by means of a post-Newtonian correction to the gravitoelectric influence of the Sun.  The general
relativistic effect of the rotation of the Sun on planetary orbits was first calculated by de Sitter [3] and later more generally by Thirring
and Lense [4].  Evidence for the Lense-Thirring precession of laser-ranged satellites LAGEOS and LAGEOS II around the rotating Earth has been
described by Ciufolini [5].  The rotation of the Earth generates a dipolar gravitomagnetic field; a direct measurement of
this field via the precession of gyroscopes in orbit about the Earth is one of the main objectives of NASA's GP-B that will
be launched in the next few years [6].

The Lense-Thirring orbital precession and the gravitomagnetic gyroscope precession point to a certain temporal
structure around a rotating mass.  This is best expressed via the gravitomagnetic clock effect [7].  In it simplest
form, the clock effect involves the difference in the periods of two clocks moving in opposite directions on the same
orbit.  Let $\tau_+ (\tau_-)$ be the proper period of a standard clock that makes a complete revolution around a rotating
mass on a prograde (retrograde) orbit; then, for a circular orbit of radius $r \gg 2 GM/c^2$ in the equatorial plane,
$\tau_+ - \tau_- \approx 4\pi J/Mc^2$.  The gravitomagnetic clock effect is in the lowest order independent of Newton's
gravitational constant and the radius of the orbit.  The key requirement for the existence of this effect is 
\emph{azimuthal closure} [8].  For a circular equatorial orbit around the Earth, $\tau_+ - \tau_- \approx 10^{-7}$ sec;
however, the detection of this effect would involve a difficult experiment [9]. Various complications have been discussed in
recent studies [10].  An important feature of the gravitomagnetic clock effect is that the prograde period is $\it longer$
that the retrograde period; this is contrary to what would be expected
on the basis of the heuristic notion of ``dragging''
associated with rotating masses [8].

There are different approaches to GEM within the framework of general relativity [11].  For the sake of simplicity and
convenience, we adopt a convention that results in as close a connection with the formulas of the standard electrodynamics as
possible [12].  In this convention, a test particle of inertial mass $m$ has gravitoelectric charge $q_E = -m$ and
gravitomagnetic charge $q_B = -2m$.  If the source is a rotating body of inertial mass $M$, the corresponding gravitational
charges are positive, i.e. $Q_E =M$ and $Q_B=2M$, in order to ensure that gravity is attractive.  Thus, we always have
$q_B/q_E = 2$; this can be traced back to the spin-2 character of the gravitational field.  Hence for a spin-1 field
$q_B/q_E = 1$, as in Maxwell's theory.

\section{GEM Field Equations}

Let us begin with a linear perturbation of Minkowski spacetime given by $g_{\mu \nu} = \eta_{\mu \nu} + h_{\mu \nu}$, where
$x^{\mu} = (ct, {\bf{x}})$.  It is useful to define the quantity $\bar{h}_{\mu\nu} = h_{\mu\nu}-\frac {1}{2}\eta_{\mu\nu}h$,
where $h=$tr$(h_{\mu\nu})$.  The gravitational field equations then assume the form
\begin{equation}
\Box \bar{h}_{\mu\nu} = -\frac{16 \pi G}{c^4} T_{\mu\nu}
\end{equation}
once the Lorentz gauge condition $\bar{h}^{\mu\nu}, _{\nu} = 0$ is imposed.  We are interested in the special
retarded solution of (1) given by
\begin{equation}
\bar{h}_{\mu\nu} = \frac {4G}{c^4} \int \frac {T_{\mu\nu}(ct-|{\bf x}-{\bf x^{\prime}}|, {\bf x^{\prime}})} {|{\bf x}-{\bf x^{\prime}}|}\;
d^{3}{\bf x^{\prime}}\;.
\end{equation}
Setting $T^{00} = \rho c^{2}$ and $T^{0i} = cj^{i}$, where $\rho$ is the matter density and $\bf{j}$ is the matter
current, inspection of solution (2) reveals that  $\bar{h}_{00}= 4\Phi/c^{2}, \bar{h}_{0i} =-2A_{i}/c^{2}$, and
$\bar{h}_{ij} = O(c^{-4})$.  Here $\Phi$ and $\bf A$ have the interpretation of GEM scalar and vector potentials,
respectively.  Far from the source, e.g., $\Phi\sim GM/r$ and ${\bf{A}} \sim G{\bf{J}} \times {\bf{x}}/c\,r^{3}$; in
general, $\Phi$ reduces to the Newtonian gravitational potential while ${\bf{A}} = O(c^{-1})$.  Neglecting terms of
$O(c^{-4})$, we find that the Lorentz gauge condition reduces to
\begin{equation}
\frac {1}{c} \frac {\partial \Phi}{\partial t} + {\bf \nabla} \cdot (\frac{1}{2} {\bf{A}}) = 0.
\end{equation}
The spacetime metric is then of the form
\begin{equation}
ds^{2} = -c^{2} (1- 2 \frac{\Phi}{c^{2}}) dt^{2} - \frac {4}{c} ({\bf{A}} \cdot d {\bf{x}}) dt + (1 + 2 \frac {\Phi}{c^{2}})
\delta_{ij}dx^{i}dx^{j}.
\end{equation} 

The GEM fields are defined by
\begin{equation}
{\bf{E}} = -\nabla \Phi - \frac {1}{c} \frac {\partial}{\partial t} (\frac {1}{2} {\bf{A}})     \; ,\;     {\bf{B}} = {\bf
{\nabla}} \times {\bf{A}},
\end{equation}
which immediately imply the source-free GEM field equations
\begin{equation}
{\bf{\nabla}} \times {\bf{E}} = -\frac {1}{c} \frac {\partial}{\partial t} (\frac {1}{2} {\bf{B}})   \;, \; {\bf{\nabla}}
\cdot (\frac{1}{2} {\bf{B}}) = 0,
\end{equation}
while the gravitational field equations (1) imply
\begin{equation}
{\bf{\nabla}} \cdot {\bf{E}} = 4\pi G\rho    \; , \;   {\bf{\nabla}} \times (\frac {1}{2} {\bf{B}}) = \frac
{1}{c}\frac {\partial}{\partial t} {\bf{E}} + \frac {4 \pi G}{c} {\bf {j}}.
\end{equation}
These equations contain the conservation law for mass current $\partial\rho/\partial t + {\bf{\nabla}} \cdot {\bf{j}}
=0$, as they should.  Moreover, the gravitomagnetic field appears in (6) -(7) with a factor of $\frac {1}{2}$  since
$q_{B}/q_{E} = 2$. 

Let us recall that the gravitational potentials $h_{\mu\nu}$ are gauge dependent; that is, under a coordinate transformation
$x^{\mu}\rightarrow x^{\prime \mu} = x^{\mu} - \epsilon^{\mu}, h_{\mu\nu} \rightarrow h^{\prime}_{\mu\nu}= h_{\mu\nu} +
\epsilon_{\mu,\nu} +\epsilon_{\nu,\mu}$ to linear order.  Therefore,
\begin{equation}
\bar{h}^{\prime}_{\mu\nu} = \bar{h}_{\mu\nu} + \epsilon_{\mu,\nu} + \epsilon_{\nu,\mu} -\eta_{\mu\nu}\epsilon^{\alpha},_{\alpha}
\end{equation}
and the Lorentz gauge condition is preserved provided $\Box \epsilon_{\mu} = 0$.  In general, the gauge freedom
leaves the curvature tensor invariant but not the connection; however, it is possible to restrict the coordinate freedom in
such a way that the GEM fields remain invariant.  To this end, let $\epsilon ^{0} = O(c^{-3})$ and $\epsilon^{i}
=O(c^{-4})$; then, $\epsilon^{\alpha},_{\alpha} = O(c^{-4})$ and with $\Psi=c^{2}\epsilon^{0}/4$ we find that GEM
potentials are defined up to the gauge transformation
\begin{equation}
\Phi^{\prime} = \Phi - \frac {1}{c} \frac{\partial}{\partial t}\Psi   \; ,\;   {\bf{A}}^{\prime} = {\bf{A}} + 2  {\bf \nabla} \Psi,
\end{equation}
where $\Psi$ is a solution of $\Box \Psi =0$ (``Lorentz gauge").  The GEM fields, which are in fact elements of the
Christoffel connection, remain invariant under the gauge transformation in this approximation scheme.

The motion of a test particle in the GEM field follows from the Lagrangian $L= -m\; c\; ds /dt$ that can be evaluated to linear
order in $\Phi$ and $\bf{A}$ from (4).  The result is
\begin{equation}
L= -mc^{2} ( 1 - \frac{v^2}{c^2})^{\frac{1}{2}} + m\gamma(1+ \frac{v^2}{c^2}) \Phi - \frac{2m}{c}\gamma {\bf v} \cdot {\bf
A},
\end{equation}
where $\gamma$ is the Lorentz factor.  If $\partial{\bf A}/ \partial t = 0$, we find that to lowest order in ${\bf v}/c$,
\begin{equation}
{\bf F} = q_{E} {\bf E} + q_{B} \frac {\bf v}{c} \times {\bf B}
\end{equation}
in analogy with the Lorentz force law.  This follows from the fact that the canonical momentum of a test particle
is given by ${\bf p} + (-2m/c){\bf A}$, where $\bf p$ is the kinetic momentum.  The exchange of energy and momentum between
the particles and the GEM field could naturally lead to the construction of the Maxwell stress-energy tensor for the GEM
field.

For the sake of convenience, it is possible to express the GEM potentials using the 4-vector notation as
\begin{equation}
{\cal{A}}^{\mu} = (2 \Phi, {\bf A})
\end{equation}
so that the Lorentz gauge (3) can be written as $ {\cal A}^{\mu},_{\mu} =0$.  We set $c=1$ in the rest of this
paper, except where indicated otherwise, and define the GEM Faraday tensor ${\cal F}_{\mu \nu}$ as $2{\cal F}_{\mu \nu} =
{\cal A}_{\nu,\mu} - {\cal A}_{\mu,\nu}$.  Then the GEM Maxwell equations are ${\cal F}_{[\mu \nu,\sigma]} = 0$ and
${\cal F}^{\mu \nu},_{\nu} = 4 \pi G j^{\mu}$, where $j^{\mu} = (\rho, {\bf j}) $ is the mass current.  It is then tempting to
construct the analog of the Maxwell tensor as in electrodynamics,
\begin{equation}
G\ {\cal T}^{\alpha \beta} = \frac {1}{4 \pi} ({\cal F}^{\alpha}{}_\gamma {\cal F}^{\beta \gamma} - \frac{1}{4} \eta^{\alpha
\beta} {\cal F}_{\gamma \delta} {\cal F}^ {\gamma \delta}).
\end{equation}
However, the physical significance of this quantity is doubtful as first pointed out by Maxwell [13] in his
fundamental work on the dynamical theory of the electromagnetic field.  The basis for Maxwell's considerations was the
notion that the attractive nature of gravity would lead to a negative energy density for the field, while the
electromagnetic analogy would imply a positive result.  In fact, in our approximation scheme we can use instead the standard
pseudotensor $t_{\mu \nu}$ of general relativity that gives a negative energy density.  This is the subject of section 4.

Finally, we digress here to point out the usefulness of the 4-vector notation introduced in (12) by considering the
propagation of {\it test} electromagnetic radiation in the GEM field.  In the absence of $h_{\mu\nu}$ and in the background
inertial spacetime, let us introduce the Kramers vector ${\bf f}={\bf e} + i{\bf b}$, where ${\bf e}$ and ${\bf b}$ are the
electric and magnetic fields, respectively.  Then Maxwell's equations for the radiation field can be expressed as $
{\bf \nabla} \cdot {\bf f} = 0$ and ${\bf \nabla} \times {\bf f} = i\; \partial{\bf f}/ \partial t$.  The latter equations
can be written in the "Dirac" form using $p_{\alpha} = -i\hbar \partial_{\alpha}$. If we now consider test electromagnetic
radiation propagating in the GEM field by the standard substitution $p_{\alpha} \rightarrow p_{\alpha} - q{\cal A}_{\alpha}$,
where $q_{E} = -\hbar \omega$ and $ q_{B} = -2 \hbar \omega$ for radiation of frequency $\omega$, then we recover the wave equation
\begin{equation}
({\bf \nabla} + 2 i\omega {\bf A}) \times {\bf f} = \omega N{\bf f},
\end{equation}
where $N=1+2\Phi$ has the interpretation of the index of refraction [14].  It is interesting to note that in the
Lagrangian (10) as $ v \rightarrow c$ the coefficient of $\Phi$ becomes {\it twice the energy of the particle}, which is
also the coefficient of the term $-{\bf v} \cdot {\bf A}$ in (10); this is consistent with the relativistic substitution for
the 4-momentum $p_{\alpha} \rightarrow p_{\alpha} - q{\cal A}_{\alpha}$ with ${\cal A}_{\alpha}$ defined by (12).

\section{Gravitational Larmor Theorem}

In electrodynamics, the Larmor theorem has played an important role in the description of the motion of charged spinning
particles. Explicitly, for slowly varying fields and to linear order in ${\bf v}/c$ and field strength, the electromagnetic
field can be replaced by an accelerated system with translational acceleration ${\bf a}_{L} = -q {\bf E}/m$ and rotational
frequency ${\bf \omega}_{L} = q{\bf B}/(2mc)$.  For all charged particles with the same $q/m$, the electromagnetic forces
are locally the same as inertial forces; this is reminiscent of the principle of equivalence.  However, $q/m$ is not the
same for all charged particles; hence, a geometric theory of electromagnetism analogous to general relativity is
impossible.  The geometric treatment of gravitation is a direct consequence of the universality of the gravitational
interaction.

The gravitational Larmor theorem expresses Einstein's principle of equivalence within the GEM framework.  To see this
clearly let us consider an event in spacetime with GEM fields $\bf E$ and $\bf B$.  The corresponding Larmor quantities
would be ${\bf a}_{L} = -q_{E} {\bf E}/m={\bf E}$ and ${\bf \omega}_{L} = q_{B} {\bf B} /(2mc)= -{\bf B}/c$, in accordance
with our convention.  In the neighborhood of this event, we can replace the GEM field by a neighborhood in Minkowski
spacetime that has translational acceleration ${\bf a}_{L} = {\bf E}$ and rotational frequency ${\bf \omega} _{L} = -{\bf B}\;
($ with $\ c = 1)$.  The geodesic coordinates in the neighborhood of such an accelerated system would be $ X^{\mu} = (\tau, {\bf
X})$ such that the Minkowski metric can be written as $ds^{2} =g_{\mu\nu}dX^{\mu}dX^{\nu}$ with
\begin{equation}
g_{00} = -(1+{\bf a}_{L} \cdot {\bf X})^{2} + ({\bf \omega}_{L} \times {\bf X})^{2},\\ 
\end{equation}

\begin{equation}
g_{0i} = ({\bf \omega}_{L} \times {\bf X})_{i} \;,\;  g_{ij} = \delta_{ij}.
\end{equation}
Let us note that to linear order in {\bf X}, the metric (15)-(16) would become the same as (4) once we set $\Phi =
-{\bf a}_{L} \cdot {\bf X}$ and ${\bf A} = - \frac {1}{2} {\bf \omega}_{L} \times {\bf X}$ as in section 2 and neglect spatial
curvature.  Then the GEM fields corresponding to these potentials are ${\bf E} = -{\bf \nabla} \Phi = {\bf a} _{L}$ and
${\bf B} = {\bf \nabla} \times {\bf A} = -{\bf \omega}_{L}$, thus confirming the validity of the gravitational Larmor
theorem in our approach.

Einstein's heuristic principle of equivalence is traditionally stated in terms of the translational acceleration of the``Einstein elevator";
however, it is clear from the analysis of this section that a Larmor rotation of the elevator is necessary as well to take due account of the
gravitomagnetic field.\\

\section{Stress-Energy Pseudotensor}

The standard Landau-Lifshitz pseudotensor $t_{\mu\nu}$ is useful in a global sense for asymptotically flat spacetimes.  It
is therefore appropriate in our linear perturbation analysis around Minkowski spacetime for $r \gg 2GM/c^{2}$; however, we
use $t_{\mu\nu}$ to compute the {\it local} stress-energy content of the GEM fields.  We find that
\begin{eqnarray}
16 \pi G\ t^{00}& = & -14 E^{2} + \sum_{i,j}(A_{i,j}+A_{j,i})(A_{i,j} +A_{j,i}) - 6(\Phi,_{0})^{2} \nonumber\\
                &   &-14 E^{i} A_{i,0} -\frac {7}{2} A^{i}{}_{,0} A_{i,0},
\end{eqnarray}

\begin{eqnarray}
4 \pi G\ t^{0i} & = &\Big(-2\; {\bf E} \times {\bf B} - {\bf A},_{0} \times {\bf B} - 3\Phi,_{0} {\bf E} -\frac{3}{2} \Phi,_{0} {\bf
A},_{0}\Big)^{i},
\end{eqnarray}

\begin{eqnarray}
16 \pi G\ t_{ij} & = & 4 (E_{i}E_{j} -\frac {1}{2} \delta_{ij}E^{2}) + 4 (B_{i}B_{j} + \frac{1}{2} \delta_{ij}B^{2}) \nonumber\\
   &  &-6[E_{i}A_{j,0} + E_{j}A_{i,0} + \delta_{ij}(\Phi,_{0})^{2} - \delta_{ij} E^{k}A_{k,0}] \nonumber\\
  &  &-7(A_{i,0}A_{j,0} + \frac{1}{2} \delta_{ij}A^{k}{}_{,0}A_{k,0}).
\end{eqnarray}

For a stationary source with angular momentum ${\bf J} = J\hat{\bf z}$, the dominant contributions are
\begin{equation} 
t^{00} = - \frac {7}{8\pi} \frac{GM^{2}} {r^{4}}, (t^{0i}) = - \frac {1}{2\pi} \frac {GMJ}{r^{5}}\; \sin\theta \; \hat
{\bf \phi}, \\ 
\end{equation}

\begin{equation}
t^{ij} = \frac {1}{4\pi} \frac{GM^{2}}{r^{6}} x^{i}x^{j}.
\end{equation}
It is important to note that the field energy density is negative and that there is a flux of energy that
circulates around the mass in a direction opposite to its sense of rotation.  The flow velocity, $v_{g}^{i} = t^{0i}/t^{00}$, is
given by 
\begin{equation}
{\bf v}_{g} = \frac{4}{7} \frac {J}{Mr} \; \sin \theta \; \hat{\bf \phi}.
\end{equation}
For the sake of comparison, it is interesting to point out that the Maxwell tensor (13) would give the same
result as (22) except that the numerical coefficient would be $1$ instead of $\frac {4}{7}$.\\

\section{GEM Stress-Energy Tensor}

Thus far our perturbative treatment of GEM has involved a fixed global background inertial frame.  This is contrary to the
spirit of the general theory of relativity; therefore, it is important to develop an alternative approach for an arbitrary
gravitational field.  To this end, imagine a congruence of test geodesic observers in a gravitational field.  Choosing an
observer in the congruence as a reference observer, one can set up a Fermi coordinate system in its neighborhood.  Instead
of a global background inertial frame, we thus have a local inertial frame along the worldline of the fiducial observer. 
The spacetime metric in the Fermi system is given by
\begin{eqnarray}
g_{00} & = &- 1 - R _{0i0j} (\tau) X^{i}X^{j} +...,  \\
g_{0i} & = & - \frac{2}{3} R_{0jik} (\tau) X^{j}X^{k} +...,  \\
g_{ij} & = &\delta_{ij} - \frac {1}{3} R_{ikjl} X^{k}X^{l} +...,
\end{eqnarray}
where $X^{\mu} = (\tau, {\bf X})$ are Fermi coordinates and
\begin{equation}
R_{\alpha \beta \gamma \delta} = R_{\mu \nu \rho \sigma} \lambda^{\mu}_{(\alpha)} \lambda^{\nu}_{(\beta)}
\lambda^{\rho}_{(\gamma)} \lambda^{\sigma}_{(\delta)}
\end{equation}
are the components of the Riemann tensor projected on the Fermi-propagated orthonormal tetrad frame
$\lambda^{\mu}_{(\alpha)}(\tau)$ of the fiducial observer at the origin of spatial Fermi coordinates $(\tau, {\bf 0})$.  Let
us note that the deviations from the local fiducial inertial frame are of the form
\begin{eqnarray}
\Phi (\tau, {\bf X}) & = &- \frac {1}{2} R_{0i0j} (\tau) X^{i}X^{j} +...,  \\
A_{i} (\tau, {\bf X})& = &\frac {1}{3} R_{0jik} (\tau) X^{j}X^{k} +...,
\end{eqnarray}
where in this GEM approach we simply ignore the spatial curvature.  The corresponding GEM fields would then be
\begin{eqnarray}
E_{i}(\tau, {\bf X}) & = &R_{0i0j} (\tau) X^{j} +..., \\
B_{i}(\tau, {\bf X}) & = &-\frac {1}{2} \epsilon_{ijk} R_{jk0l} (\tau) X^{l} +...,
\end{eqnarray}
where we have used our previous conventions.

From the viewpoint of our reference observer, the other test observers in the congruence move in accordance with the
generalized Jacobi equation [15]\nonumber \\

\begin{eqnarray}
\frac {d^{2} X^{i}}{d \tau^{2}} + R_{0i0j}X^{j}+ 2 R_{ikj0} V^{k}X^{j} + (2 R_{0kj0}V^{i}V^{k} \nonumber \\
 + \frac {2}{3} R_{ikjl} V^{k}V^{l} + \frac {2}{3} R_{0kjl} V^{i}V^{k}V^{l}) X^{j} = 0,
\end{eqnarray}
which is valid to first order in the separation $\bf X$ while the rate of separation ${\bf V} = d{\bf X}/d\tau$ could be
arbitrary $(|{\bf V}|<1)$.  Restricting the deviation equation (31) to first order in the relative velocity as well, we find
that it can be written in the form
\begin{equation}
m \frac {d^{2}{\bf X}}{d \tau^{2}} = q_{E}{\bf E} + q_{B}{\bf V} \times {\bf B},
\end{equation}
with $q_{E} = -m$ and $q_{B} = -2 m$ as before.  It is interesting to note that the gravitoelectric and gravitomagnetic
fields (29)-(30) are given by the electric and magnetic components of the Riemann tensor, respectively.

It is possible to combine (29) and (30) and define a GEM Faraday tensor by $F_{\alpha\beta} = - R_{\alpha\beta 0i}X^{i}$ to
linear order in $\bf X$.  Here $F_{0i} = -E_{i}$ and $F_{ij} = \epsilon_{ijk} B_{k}$ as in standard electrodynamics; therefore,
this treatment is in general different from our perturbative approach.  This Faraday tensor satisfies Maxwell's equations
$F_{[\alpha \beta,\gamma]} =0$ and $F^{\alpha \beta},_{\beta} = 4 \pi J^{\alpha}$ with $4 \pi J_{\alpha} (\tau,{\bf 0}) =
-R_{0 \alpha}$ along the fiducial worldline.  Therefore, the classical field theory of the GEM field in the Fermi frame can
be developed in the standard manner [16].  We are particularly interested in the Maxwell stress-energy tensor $T^{\alpha
\beta}$ associated with $F_{\alpha\beta}$; this is given, in a similar form as (13), by
\begin{equation}
G\ T^{\alpha \beta}(\tau, {\bf X}) = \frac {1}{4 \pi} (R^{\alpha}{}_{\gamma 0i} R^{\beta \gamma} {}_{0j} - \frac {1}{4}\eta ^{\alpha
\beta} R_{\gamma \delta 0i} R^{\gamma \delta} {}_{0j}) X^{i} X^{j}.
\end{equation}
The fact that $T^{\alpha \beta}$ vanishes for the reference observer is consistent with Einstein's principle of
equivalence.  Away from this fiducial observer, $T^{\alpha \beta}$ is nonzero in general;  but, we could have just as well
chosen a different reference observer in the congruence and then the situation would have been reversed.  In addition, we
recall that physical measurements of field energy and momentum cannot be performed at a point and require an averaging
process as already emphasized by Bohr and Rosenfeld [17].  Therefore, we average (33) at an event $(\tau,{\bf 0})$ along the
reference trajectory over an infinitesimal sphere of radius $\epsilon \ell$.  Here $0<\epsilon\ll 1$ and $\ell$ is an
invariant length scale (e.g., $\ell$ could be $GM/c^{2}$ or the Planck length when a natural length scale is absent in the
case under consideration). Then $< X^{i}X^{j}> = C_{0}(\epsilon \ell)^{2} \delta_{ij}$, where $C_{0} = \frac{1}{5}$ or
$\frac {1}{3}$ for averaging over the volume or the surface of the sphere, respectively; in any case, this constant
coefficient can always be absorbed in the definition of $\ell$.  Hence
\begin{equation}
<T_{\alpha\beta}> = \frac{C_{0} \epsilon ^{2} \ell^{2}}{4\pi G} \tilde{T}_{\mu\nu\rho\sigma} \lambda^{\mu}_{(\alpha)} \lambda
^{\nu}_{(\beta)} \lambda ^{\rho}_{(0)} \lambda^{\sigma}_{(0)},
\end{equation} 
where $\tilde{T}_{\mu\nu\rho\sigma} (x)$ is the Bel tensor given by
\begin{equation}
\tilde{T}_{\mu\nu\rho\sigma} (x) = \frac{1}{2} (R_{\mu \xi \rho \zeta} R_{\nu}{}^{\xi}{}_{\sigma}{}^{\zeta} + R_{\mu\xi\sigma\zeta}
R_{\nu}{}^{\xi}{}_{\rho}{}^{\zeta}) - \frac{1}{4} g_{\mu\nu} R_{\alpha\beta\rho\gamma} R^{\alpha \beta} {}_{\sigma} {}^{\gamma}.
\end{equation}
This tensor was first defined by Bel in 1958 for Einstein spaces on the basis of a certain analogy with the
Maxwell stress-energy tensor [18].  It is symmetric in its first and second pair of indices and traceless in the first
pair.  We define the GEM stress-energy tensor to be $\tilde {T}_{\mu\nu} (x)$, where the stress-energy measured by an
arbitrary observer is essentially given by the projection of the Bel tensor on the tetrad of the observer, i.e.
\begin{equation}
\tilde{T}_{(\alpha)(\beta)} = \frac {\ell^{2}}{G} \;\tilde{T}_{\mu\nu\rho\sigma} \lambda^{\mu}_{(\alpha)} \lambda
^{\nu}_{(\beta)} \lambda ^{\rho}_{(0)} \lambda^{\sigma}_{(0)}\;.
\end{equation}

For a Ricci-flat spacetime, the curvature tensor reduces to the Weyl conformal tensor $C_{\mu\nu\rho\sigma}$ and the Bel
tensor reduces to the totally symmetric and traceless Bel-Robinson tensor $T_{\mu\nu\rho\sigma}$ given by
\begin{equation}
T_{\mu\nu\rho\sigma} = \frac{1}{2} (C_{\mu \xi \rho \zeta} C_{\nu}{}^{\xi}{}_{\sigma}{}^{\zeta} + C_{\mu\xi\sigma\zeta}C_{\nu}{}^{\xi}{}_{\rho}{}^{\zeta})- \frac{1}{16} g_{\mu \nu} g_{\rho \sigma} C_{\alpha\beta\gamma\delta} C^{\alpha\beta\gamma\delta}.
\end{equation}
In this case, the spatial components of the curvature tensor (that were ignored in our construction of the GEM
stress-energy tensor) are in fact given (up to a sign) by its electric components.  Therefore,  $\tilde{T}_{\mu\nu}(x)$ reduces
in this case to the {\it gravitational stress-energy tensor} $T^{\ast}{}_{\mu\nu}$ given by
\begin{equation}
T^{\ast}{}_{(\alpha)(\beta)} = \frac {\ell ^{2}}{G}\; T_{\mu\nu\rho\sigma} \lambda^{\mu}_{(\alpha)} \lambda
^{\nu}_{(\beta)} \lambda ^{\rho}_{(0)} \lambda^{\sigma}_{(0)}.
\end{equation}

This GEM derivation of the physical content of the Bel and Bel-Robinson tensors has made it possible to provide new
definitions for the average gravitational stresses (such as the pressure of gravitational radiation) up to a positive
multiplicative factor [16].\\

\section{Gravitational Poynting Flux}

The GEM approach based on the curvature tensor can be used in arbitrary gravitational fields.  A simple application of the
gravitational stress-energy tensor to the exterior field of a rotating source implies the existence of a steady flow of
gravitational energy circling the source.  This situation is analogous to the electromagnetic case of a charged rotating
mass; that is, there is a steady Poynting flux of electromagnetic energy circulating around the mass.  In connection with the
physical interpretation of the stress-energy tensor, Pauli has noted that in a purely electrostatic (or, alternatively,
magnetostatic) field, there is no momentum density, but there is momentum current [19], while Feynman has discussed some
counter-intuitive features of the Poynting flux [20].  Nevertheless, one can study the general physical characteristics of
this gravitational energy flow using the GEM stress-energy tensor [16].

For the exterior of a rotating mass $(r\gg2GM/c^{2})$, the dominant contributions to the Bel-Robinson tensor are
\begin{equation}
T^{(0)(0)} = \frac {3\ell^{2}GM^{2}}{r^{6}},\\ T^{(0)(i)} = \frac {9\ell^{2}GMJ}{r^{7}} ({\bf \hat{J}} \times {\bf \hat{x}})^{i},
\end{equation}

\begin{equation}
T^{(i)(j)} = \frac {\ell^{2}GM^{2}}{r^{6}} (2\delta_{ij} - 3 \hat{x}^{i} \hat{x}^{j}),
\end{equation}
for freely falling observers.

The three approaches to the stress-energy content of GEM have this feature in common: They all predict that the
gravitational energy will steadily flow around a rotating mass with a velocity
\begin{equation}
{\bf v}_{g} = \lambda \frac {J}{Mr} \;\sin \theta \; \hat{\bf \phi},
\end{equation}
where $\lambda$ is a numerical factor equal to 1, $\frac{4}{7}$, or 3 for the three cases we have discussed.  It
is therefore interesting to examine certain general features of this flow velocity.  We find that ${\bf \nabla}
\cdot {\bf v}_{g} =0$, so the flow is divergence-free; in fact, the streamlines are circles in planes perpendicular to $\bf
J$.  Moreover, the vorticity of the flow, ${\bf \omega}_{g} = {\bf
\nabla} \times {\bf v}_{g} = 2 \lambda (J/Mr^{2})\cos \theta \,{\bf \hat{r}}$, vanishes in the equatorial plane and is maximum along the rotation axis.  Let us note that $
{\bf v}_{g} = {\bf \nabla} \times {\bf \Psi}_{g}$, where ${\bf \Psi}_{g}$ is the gravitational stream function
given by ${\bf \Psi}_{g} = \lambda(J/M) \cos \theta \,{\bf \hat{r}}$.  In fact ${\bf \omega}_{g} = -{\nabla}^{2} {\bf
\Psi}_{g}$, since $\nabla \cdot {\bf \Psi}_{g} = 0$. The circulation of ${\bf v}_{g}$,
\begin{equation}
C_{g} = \oint {\bf v}_{g} \cdot d{\bf l},
\end{equation}
around a streamline is given by $ 2 \pi \lambda (J/M)$ sin$^{2} \theta$; therefore, the circulation is independent
of the radial distance $r$ and vanishes on the axis of rotation.  In the equatorial plane, the circulation $C_{g}$ is
simply proportional to the specific angular momentum of the source, just like the gravitomagnetic clock effect.  It would be
interesting to explore the observational aspects of the steady gravitational energy flux around rotating masses.

\section*{Acknowledgments}
I would like to thank the organizers of the Spanish Relativity Meeting (EREs2000) for their kind invitation and warm
hospitality.

\end{document}